\documentclass[aps,prl,reprint,superscriptaddress]{revtex4-2}
\usepackage{graphicx}
\usepackage{textcomp}
\usepackage{gensymb}
\usepackage{amsmath}
\usepackage{hyperref}

\begin{document}
	
\title{Superconducting vortex-charge measurement using cavity electromechanics}

\author{Sudhir~Kumar~Sahu}
\affiliation{Department of Physics, Indian Institute of Science, Bangalore-560012 (India)}

\author{Supriya~Mandal}
\affiliation{Department of condensed matter physics and material sciences, 
	Tata Institute of Fundamental Research, Mumbai - 400005 (India)} 

\author{Sanat~Ghosh}
\affiliation{Department of condensed matter physics and material sciences, 
	Tata Institute of Fundamental Research, Mumbai - 400005 (India)} 

\author{Mandar~M.~Deshmukh}
\affiliation{Department of condensed matter physics and material sciences, 
	Tata Institute of Fundamental Research, Mumbai - 400005 (India)} 

\author{Vibhor~Singh}
\email[]{v.singh@iisc.ac.in}
\affiliation{Department of Physics, Indian Institute of Science, 
	Bangalore-560012 (India)}

\begin{abstract}As the magnetic field penetrates the surface of a superconductor, 
it results in the formation of flux-vortices. It has been predicted 
that the flux-vortices will have a charged vortex core and create
a dipolelike electric field. Such a  charge trapping in vortices is 
particularly enhanced in high-T$_c$ superconductors (HTS). 
Here, we integrate a mechanical resonator made of a thin flake of 
HTS Bi$_2$Sr$_2$CaCu$_2$O$_{8+\delta}$ into a 
microwave circuit to realize a cavity-electromechanical device. 
Due to the exquisite sensitivity of cavity-based devices to the external 
forces, we directly detect the charges in the flux-vortices by 
measuring the electromechanical response of the mechanical resonator. 
Our measurements reveal the strength of surface electric dipole moment
due to a single vortex core to be approximately 30~$|e|a_B$, equivalent
to a vortex charge per CuO$_2$ layer of $3.7\times 10^{-2}|e|$,
where $a_B$ is the Bohr radius and $e$ is the 
electronic charge.
\end{abstract}

\maketitle

\begin{figure*}
	\begin{center}
		\includegraphics[width = 150mm]{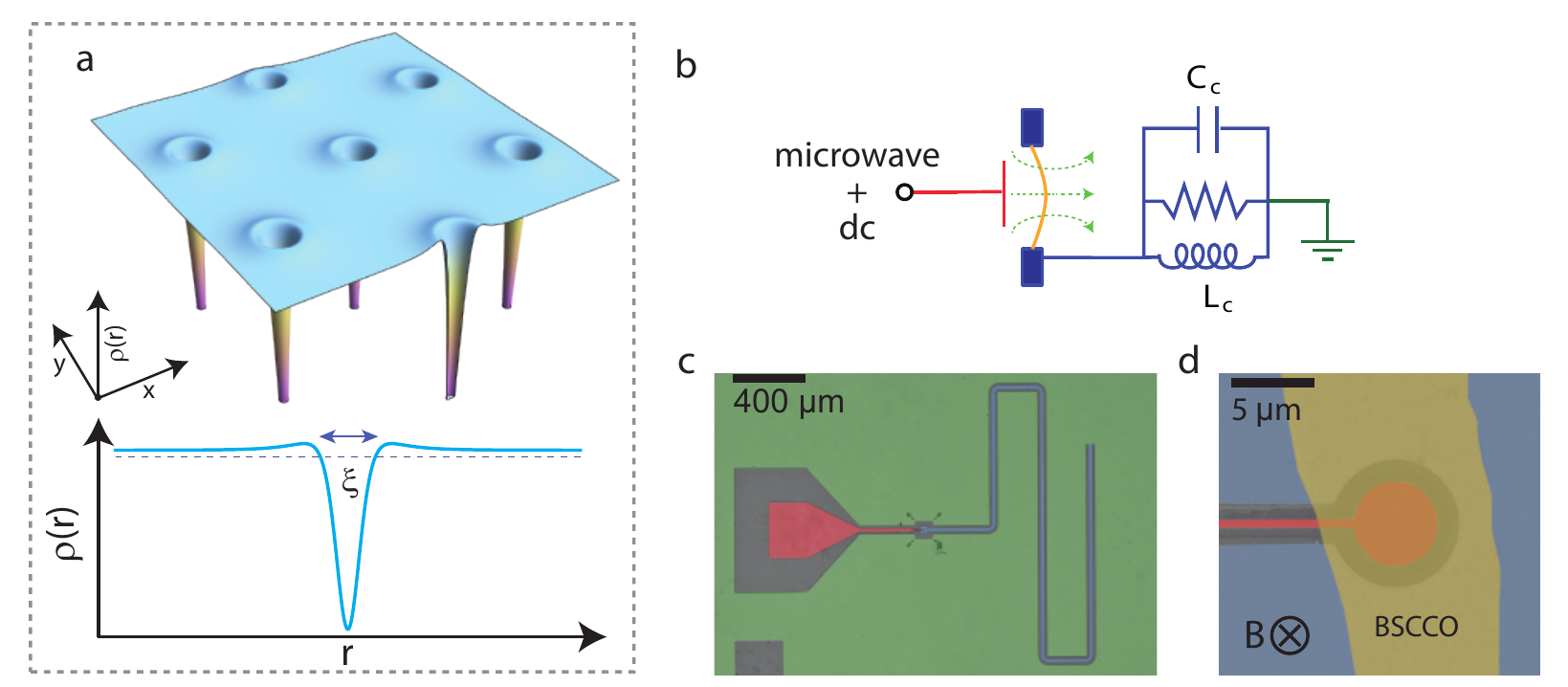}
		\caption{\textbf{Cavity optomechanical architecture for 
				detecting the vortex charge:} 
			(a) Schematic showing the redistribution of the charge density 
			at vortex core due to the chemical potential mismatch 
			in the normal and in the superconducting state.
			The cyan curve shows the radial profile of the charge density 
			redistribution around the vortex core and indicates
			the normal core of size of the 
			coherence length $\xi$.
			(b) Electromechanical system with a single-port 
			reflection cavity. The device is configured such that
			both direct current (dc) and microwave signals can be added 
			to the drive port. The equivalent lumped inductance 
			and capacitance are 1.42~nH and 350~fF, respectively.
			(c) Optical microscope image (false color) of the 
			complete device. It shows a quarter-wavelength 
			reflection-cavity in coplanar waveguide geometry 
			patterned on an intrinsic Si substrate. 
			(d) Scanning electron micrograph (false color) of 
			the device near the coupling capacitor showing the 
			suspended part of BSCCO flake which is 5 UC thick 
			($\sim$15~nm). The circular suspended part has a 
			diameter of 7~$\mu$m.
		}
		\label{fig1}
	\end{center}
\end{figure*}

The penetration of a magnetic field in type-II superconductors 
in the form of Abrikosov vortices is well-known 
\cite{ketterson_superconductivity_1999,abrikosov_nobel_2004}. 
Such vortices have a normal-core of size of the coherence length 
and each vortex is surrounded by a circulating supercurrent that decays
over a characteristic magnetic length scale. 
It has been predicted that such vortices can trap charges 
\cite{caroli_bound_1964,khomskii_charged_1995,feigelman_sign_1995,
	kolacek_charge_2001}. 
The origin of the vortex charge can be understood from the difference 
in the chemical potentials in the superconducting-state and the
normal region due to the particle-hole asymmetry. 
It leads to a redistribution of electrons to maintain
the same electrochemical potential across a vortex \cite{blatter_electrostatics_1996},
as shown schematically in Fig.~\ref{fig1}(a).
Intuitively, the idea of charged vortex core can also be captured by 
considering the inertial and Lorentz forces acting on the Cooper pairs
encircling the normal core, resulting in the depletion of charges 
from the core \cite{kolacek_charge_2001}.

To probe the trapped charge in the vortex core, the high-$T_c$ 
superconductors (HTS) are particularly attractive. 
The magnitude of the vortex charge is approximately given 
by $|e|\left(\Delta/\epsilon_F\right)^2$,
where $\Delta$ is the superconducting gap and $\epsilon_F$ is the Fermi energy.
The ratio $\Delta/\epsilon_F$ is usually high in HTS $(\Delta/\epsilon_F\sim0.1)$.
Indeed, in the past, the inference of  the charged vortex core has 
been made on bulk crystals of cuprate superconductors
based on techniques probing 
the sign reversal of Hall coefficient \cite{hagen_flux-flow_1991}, 
the nuclear quadrupole resonance \cite{kumagai_charged_2001, mitrovic_spatially_2001}, 
and instability of the vortex lattice \cite{mounce_charge-induced_2011}.
However, a measurement of the charged vortex core directly measuring its 
electrostatic field is a challenging task due to its small magnitude 
and screening by the surrounding charges.

In recent years, nanoelectromechanical systems based approach 
to probe phase transitions and the thermodynamical properties such as heat capacity, 
thermal-conductivity, and magnetization have drawn a lot of 
attention \cite{bolle_observation_1999,schwab_measurement_2000,
	chen_modulation_2016,morell_optomechanical_2019,siskins_magnetic_2020}.
Particularly, the integration of exfoliated thin flakes into
cavity optomechanical and hybrid devices have resulted into enhanced 
sensitivity 
to external forces \cite{singh_optomechanical_2014, 
weber_coupling_2014,reserbat-plantey_electromechanical_2016}. 
With these motivations in mind, we apply cavity optomechanical 
techniques to detect the charges associated with vortices in a 
high-$T_c$ superconductor by directly probing their electrostatic effect.

In this work, we use a mechanical resonator of 5 unit cell (UC) thick 
exfoliated flake of Bi$_2$Sr$_2$CaCu$_2$O$_{8+\delta}$ (BSCCO) and integrate 
it with a microwave cavity to realize a cavity electromechanical device.
Due to their low mass, such mechanical resonators have a large
coupling with the microwave field of the cavity, which enhances their
force sensitivity.
Fig.~\ref{fig1}(b) shows the electrical equivalent schematic of the device.
Such a device is analogous to a Fabry-Perot cavity with a mechanically
compliant mirror, where vibrations of the mirror parametrically 
modulate the resonant frequency of the cavity \cite{aspelmeyer_cavity_2014}.
We use a superconducting coplanar waveguide-based $\lambda/4$ 
microwave cavity as shown in Fig.~\ref{fig1}(c).
The cavity is fabricated by patterning a 200~nm thick sputtered 
molybdenum and rhenium (MoRe) alloy film on top of intrinsic silicon
wafer \cite{singh_molybdenum-rhenium_2014,sahu_nanoelectromechanical_2019}.
The MoRe film has a $T_c$ of 11~K. Near the coupler end of the 
microwave cavity, a feedline is selectively 
etched to reduce the MoRe film thickness by 120~nm.
The difference in MoRe film thickness provides the clearance to form a 
suspended mechanical drumhead resonator by the transfer of
an exfoliated BSCCO flake.
A scanning electron microscopy (SEM) image of the suspended 
BSCCO flake after the transfer is shown in Fig.~\ref{fig1}(d). 
In this configuration, the BSCCO drumhead-shaped mechanical 
resonator provides 
capacitive coupling between the feedline and microwave cavity. 
In addition, the device design allows application of both alternating current 
(ac) and direct current (dc) signals 
through the feedline.

%%%%%%%%%%%%%%%%%%%%%%%%%%%%%%%%%%%%%%%%%%%%%%%%%%%%%%%%%%%%%%%%%%%%%%%%%%%%%%%%%%%%%%%%
% Figure 2 description

\begin{figure*}
	\begin{center}
		\includegraphics[width = 150mm]{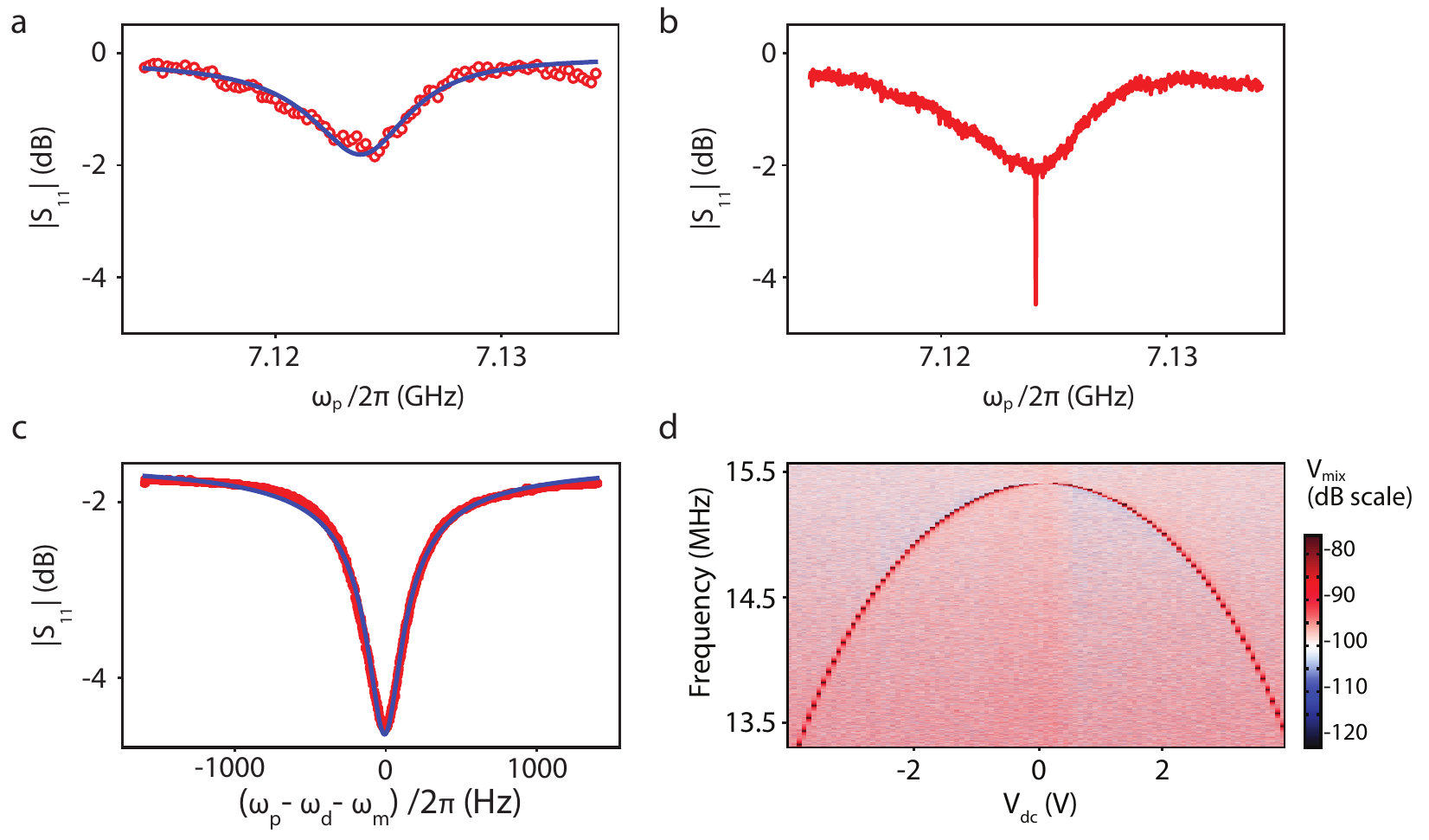}
		\caption{\textbf{Optomechanically induced absorption (OMIA) 
				measurements: } 
			(a) Normalized reflection $|S_{11}(\omega)|$ measurement of cavity 
			at 20~mK. 
			The blue line shows the fitted curve.
			(b) Measurement of $|S_{11}(\omega)|$ in the
			presence of a red-detuned pump signal $\omega_d = 
			\omega_c-\omega_m$. 
			The sharp feature at the 
			center of the cavity shows the optomechanically induced absorption
			(OMIA), resulting from the coherent oscillation of the mechanical 
			resonator.
			(c) Zoomed-in view of the OMIA feature showing the detailed 
			mechanical response. 
			(d) Colorplot of the demodulated signal $V_{mix}$ showing the 
			quadratic tuning of the mechanical resonant frequency with 
			$V_{dc}$ applied at the feedline.
		}
		\label{fig2} 
	\end{center}
\end{figure*}

The device is cooled down to 20~mK in a dilution refrigerator 
with sufficiently attenuated microwave input lines. 
Additional details of the low-temperature setup are included 
in the Supporting Information.
The reflection from the cavity $|S_{11}(\omega)|$ is measured using a 
vector network analyzer.
Fig.~\ref{fig2}(a) shows the measurement of $|S_{11}(\omega)|$ with 
fundamental resonant mode at $\omega_c/2\pi=$~7.124~GHz. 
The solid line is the fit to the cavity response, yielding 
the internal and external linewidths
of $\kappa_i/2\pi=$~0.48~MHz and $\kappa_e/2\pi=$~4.92~MHz, respectively.

For mechanical mode characterization, we drive 
the BSCCO resonator at its fundamental resonant frequency $\omega_m$ 
using the radiation-pressure force. 
To achieve this, we add a pump signal at $\omega_d=\omega_c-\omega_m$ 
along with a weak probe signal $\omega_p$ near $\omega_c$. 
In the presence of the pump and the probe signals, the mechanical 
resonator experiences a beating radiation-pressure force near 
its resonant frequency which drives it coherently \cite{weis_optomechanically_2010,zhou_slowing_2013,singh_optomechanical_2014}.
The mechanical motion of the resonator, in turn, modulates the 
intracavity pump field, resulting in an upconverted signal exactly at the 
probe frequency.
As a coherent process, the upconverted signal interferes 
with the original probe signal and therefore results in an 
interference feature with linewidth set by the mechanical 
losses. Such interference in the cavity
reflection is referred to as optomechanically induced 
absorption (OMIA) \cite{zhou_slowing_2013, singh_optomechanical_2014}.

The OMIA interference appears as a sharp absorption 
feature in the cavity response as shown in 
Fig.~\ref{fig2}(b). 
A narrow span measurement of the absorption feature is shown in Fig.~\ref{fig2}(c).
The absorption dip directly 
manifests the coherently driven mechanical response, 
and thus allows the complete characterization of the mechanical 
resonator.
From this measurement, we determine the fundamental mechanical 
mode frequency of $\omega_m/2\pi$~$\sim$15.383~MHz and linewidth of 
$\gamma_m/2\pi$~$\sim$165~Hz, corresponding to a quality 
factor of $Q_m\sim93000$.
The high quality factor is a direct indication of the 
low contact resistance between the MoRe film and BSCCO
\cite{sahu_nanoelectromechanical_2019}.
For the cavity-electromechanical device studied here, 
we estimated the single photon coupling rate 
$g_0=\left(\partial\omega_c/\partial x\right)x_{zp}$ 
to be $2\pi\times$30~Hz, where $x_{zp}$ is the quantum zero-point 
motion of the BSCCO resonator. 

%%%%%%%%%%

Due to the novel electrical design of the device, we can 
add a dc voltage $V_{dc}$ across the capacitor formed by 
the BSCCO resonator and the MoRe feedline underneath. 
While such a scheme has a significant advantage for probing the vortex 
charge \cite{blatter_electrostatics_1996}, it allows
characterization of the BSCCO resonator by an independent technique.
A dc signal, $V_{dc}$, and an ac signal, $V_{ac}$, can be used to 
actuate the resonator by a force $C_gV_{dc}V_{ac}/z_0$, where
$z_0$ is the separation between BSCCO and the bottom electrode. 
By using the cavity as an interferometer, the mechanical mode
is detected by demodulating the reflected microwave signal 
from the cavity. 
Fig.~\ref{fig2}(d) shows the plot of demodulated signal $(V_{mix})$ as 
$V_{dc}$ 
is varied. We observe a quadratic tuning of the mechanical resonant frequency
due to capacitive spring softening \cite{kozinsky_tuning_2006}.

%%%%%%%%%%%%%%%%%%% FIG3 description

\begin{figure*}
	\begin{center}
		\includegraphics[width = 165 mm]{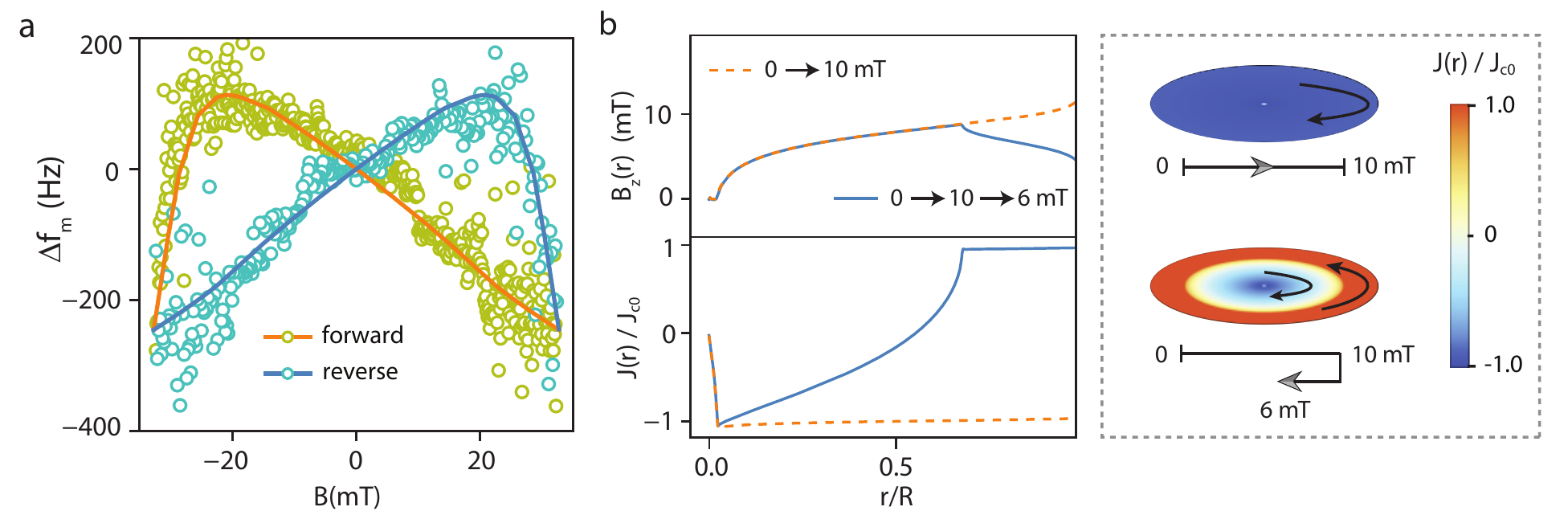}
		\caption{\textbf{Modification of the mechanical response in the 
				magnetic field:} 
			(a) Shift in the mechanical frequency with magnetic field at 
			$V_{dc}=0$~V,
			as magnetic field is swept in forward and reverse direction. 
			The solid lines show the results obtained from 
			the critical state model.
			(b) Plot of calculated local magnetic field and normalized critical 
			current computed in the BSCCO plane. Dotted lines show the spatial 
			profile after a forward sweep of $B=~$0-10~mT. Solid lines 
			show the profiles for a subsequent field descent to $B=~$6~mT.
			The side panel shows a side view of the normalized 
			current distribution for the two cases stressing upon the change in 
			the direction of current as field is lowered after reaching a 
			maximum value.
		}\label{fig3}
	\end{center}
\end{figure*}

Having described the experimental technique, we measure
the electromechanical response when a small magnetic field,
perpendicular to the CuO$_2$ plane of the BSCCO crystal is
applied. It is represented in Fig.~\ref{fig1}(d).
We perform OMIA measurements while sweeping the magnetic field 
in the forward and reverse directions.
In the presence of the magnetic field, both the cavity frequency 
and the mechanical resonant frequency can change. 
Therefore, we first record the cavity response and readjust the pump 
frequency based on the measurement of $\omega_c(B)$. 
Fig.~\ref{fig3}(a) shows the plot of the mechanical resonant 
frequency shift $\Delta f_m(B) = f_m(B) - f_m(0)$ at $V_{dc}=0$~V.
We limit the measurements in the range of $\pm$32.6~mT due to 
the reduction in the cavity quality factor at higher fields, 
which obscure the OMIA feature.

%%%%

The hysteresis in $\Delta f_m (B)$ suggests the role of 
flux-pinning, the Lorentz force of vortex transferring 
to the lattice and hence the irreversibility of the 
frequency shift, similar to the 
magnetostriction measurements on bulk 
single crystal of BSCCO \cite{ikuta_giant_1993}.
For quantitative analysis, we use the critical state model, 
which is widely used to explain the magnetization of HTS 
\cite{bean_magnetization_1962,brandt_superconductors_1996}.
Here the maximum sheet current is capped at a critical current 
which in general could be magnetic field dependent \cite{chen_kim_1989,senoussi_exponential_1988}.
Due to this sheet current and the 
local magnetic field $B_z(r)$, the flake experiences 
a Lorentz force in the lateral direction.
It results in a magnetic field dependent tension in the
flake and hence a change in the mechanical resonant 
frequency as observed in Fig.~\ref{fig3}(a).

We use a local critical sheet current as 
$J_c(r)=J_{c0} \exp(-|B_z(r)|/B_0)$, where $B_z(r)$ is the 
local magnetic field in the BSCCO plane, and the 
maximum critical current $J_{c0}$ and characteristic magnetic 
field $B_0$ are two model parameters
\cite{chen_kim_1989,senoussi_exponential_1988}.
We note that here BSCCO resonator thickness (5~UC) is much 
smaller than the London penetration length \cite{harshman_magnetic_1991}.
Therefore, a small field can push the flux-front, the boundary 
between the Meissner phase and the mixed phase, to nearly the center of the 
BSCCO flake. 
We assume that the steady-state critical currents 
remain confined to the boundaries of the BSCCO flake.
To keep the calculations simple, we treat BSCCO as
a thin circular plate. Calculation for the realistic 
geometry would only results in a scaling factor.  
Our model neglects geometrical \cite{zeldov_geometrical_1994} 
or Bean-Livingston barrier \cite{bean_surface_1964}.
Furthermore, we assume no interaction between the vortex lines 
passing through the MoRe film (bottom capacitor plate) and 
BSCCO resonator (top capacitor plate).

To calculate the force experienced by the suspended part of the 
BSCCO flake, we first calculate the radial profile of the sheet current 
density $J(r)$ and $B_z(r)$ \cite{shantsev_thin_1999}. 
The Lorentz force on the suspended part of the BSCCO flake is 
then estimated by $\int_{0}^{R} J(r) B_z(r) 2\pi rdr$, resulting 
in a net stress in the flake.
The stress can be compressive or tensile depending on 
the sign of the integral. The Lorentz-force induced 
stress can be used to compute the frequency shift of mechanical 
resonator, as shown by the solid lines in Fig.~\ref{fig3}(a).
The radial profiles of the local magnetic field and the 
normalized critical current are shown in Fig.~\ref{fig3}(b).
For these calculations, we have used the critical current density 
of $J_{c0}=3680~\text{A}/\mu\text{m}$ 
\cite{you_superconducting_2005,stangl_ultra-high_2021} and 
$B_0=100$~mT \cite{sunwong_angular_2011}.
Additional details of the model are included in the Supporting Information..

%%%%%%%%%%%%%%%%%%% FIG4 description

\begin{figure*}
	\begin{center}
		\includegraphics[width = 150 mm]{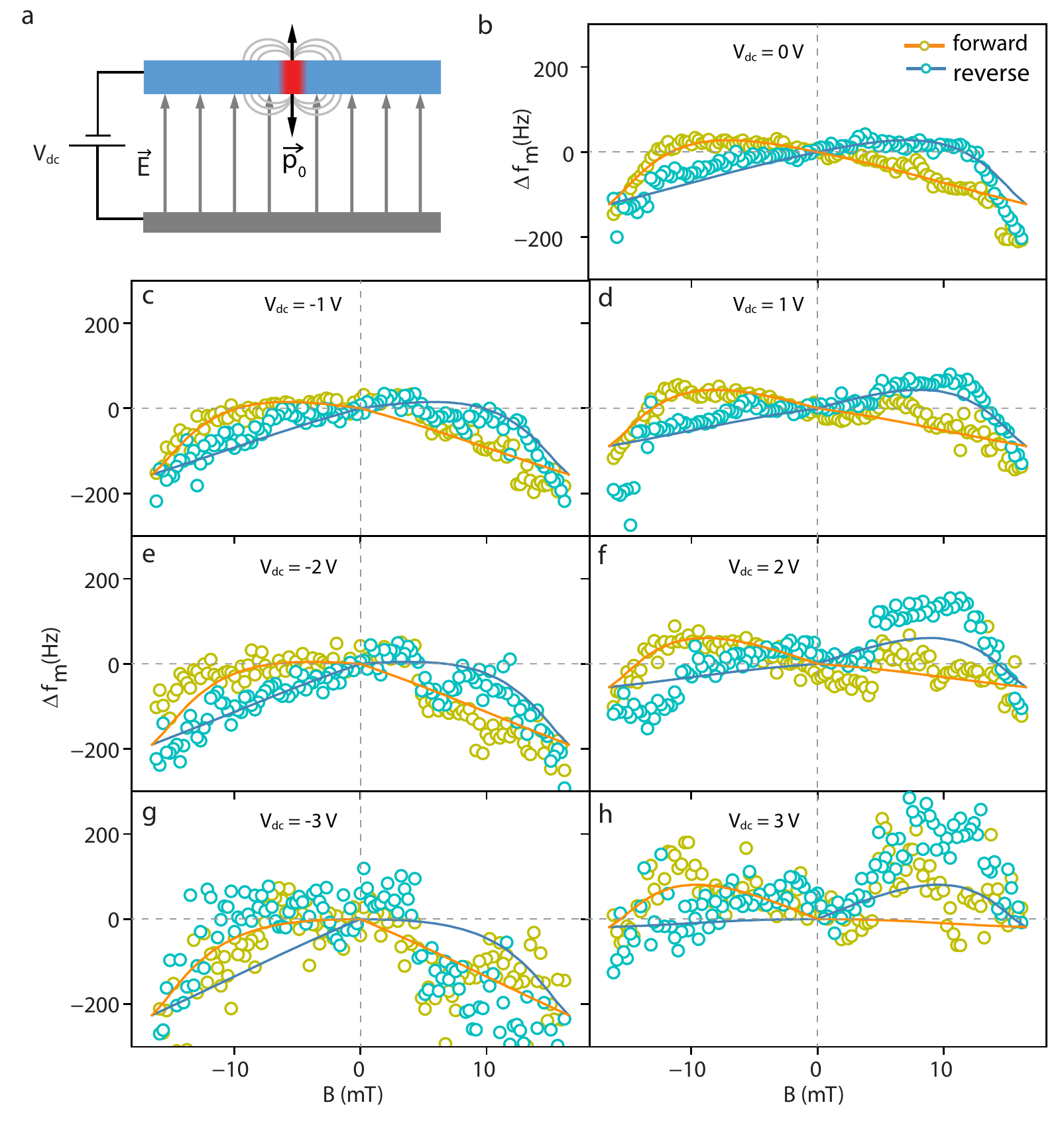}
		\caption{\textbf{Evidence for the vortex-charge: }
			(a) A schematic showing the geometry of the capacitor 
			formed by the BSCCO resonator (cyan) and MoRe plate (grey). 
			The red part of BSCCO resonator represents a single vortex, showing 
			two
			surface dipoles.
			(b-h) Measurement of shift in the mechanical frequency 
			$\Delta f_m = f_m(B, V_{dc}) - f_m(B=0, V_{dc})$
			as the magnetic field is swept in forward and reverse direction
			in the presence of a dc voltage $V_{dc}$. 
			The solid lines are the calculated curves 
			considering formation of the vortex-dipole.
		}
		\label{fig4}
	\end{center}
\end{figure*}

To probe the charge associated with vortices, their interaction 
with the electrostatic field setup by a dc voltage
plays a crucial role. 
Therefore, we now turn our focus on the mechanical response 
when $V_{dc}$ is applied  as shown schematically in 
Fig.~\ref{fig4}(a).
The plots of $\Delta f_m$ at different $V_{dc}$ applied across 
the BSCCO resonator are shown in Fig.~\ref{fig4}(b-h).
For these measurements, we sweep the magnetic field 
in $\pm$16.3~mT range due to the reduction in the mechanical 
quality factor at higher $|V_{dc}|$, which limits the measurement of 
mechanical response using OMIA technique.
The measurement of $\Delta f_m$ shows a clear asymmetry depending 
on the sign of $V_{dc}$. 
In addition, the overall dispersion with the magnetic field is enhanced 
at large negative dc voltages as compared to the case when $V_{dc}$ 
is positive.
The asymmetry in the electromechanical response with respect to the
sign of $V_{dc}$ (left vs right panels in Fig.~\ref{fig4}(b-h)
clearly suggests an electrostatic origin. 
This could be explained by considering the charge trapped
in the vortex core.
The charge per copper-oxide layer inside the vortex core is given 
by $Q_{c} \sim |e| \left(\Delta/\epsilon_F \right)^2$ \cite{khomskii_charged_1995,blatter_electrostatics_1996}.
Using $\Delta/\epsilon_F\sim0.1$, and a CuO$_2$-plane 
spacing of 0.75~nm for BSCCO \cite{zhao_sign-reversing_2019}, 
a typical estimate of $Q_{c}=0.01~|e|$ can be made. 
It is equivalent to a line charge 
density $Q_{\xi}$ of 0.013~$|e|$/nm.

In the limit $t\gg\xi$, where $t$ is the thickness of BSCCO, 
the net effect of the vortex line charge can effectively be captured 
by considering an equivalent surface electric dipole $\vec{p}_0$ 
\cite{blatter_electrostatics_1996}.
To understand our observations, we consider the interaction
of surface dipoles $\vec{p}_0$ with the electric field set up by $V_{dc}$.
A dc voltage across the capacitor modifies the interaction 
energy by $U_{dip}(z)=-\vec{p}_0\cdot\vec{E}(z)$, where $\vec{E}$ 
is the electric field between the capacitor plates, and hence leads to 
an electromechanical coupling. 
As magnetic field is increased, the number of vortices $n_v(B)$ 
penetrating the suspended part of the BSCCO increases, as
$n_v(B) = (\pi R^2 |B|)/\Phi_0$, where $\Phi_0$ is the superconducting
flux quantum and $R$ is the radius of the suspended part of BSCCO.
By considering a uniform distribution of the vortices on a triangular 
lattice across the BSCCO resonator, their contribution to the 
electromechanical energy can be calculated by summing over all the 
vortices present in the suspended part.

We include the electric dipole contribution along with the elastic, 
and the capacitive contribution to the total electromechanical energy 
$U(z)$. Therefore, the effective spring constant $\partial ^ 2U(z)/\partial z^2$ 
and hence the mechanical resonant frequency can be determined. 
The solid lines in Fig.~\ref{fig4}(b-h) show the calculated 
$\Delta f_m (B)$ by including a dc voltage dependent energy in the 
electromechanical response. Here the curves are plotted taking a dipole 
magnitude of $p_0=30~|e|a_B$ with a positively charged core, where 
$a_B = 5.29 \times 10^{-11}$~m is the Bohr radius.
It is important to point out here that the sign of the 
vortex core charge is determined by the oxygen doping level 
in the BSCCO crystal. For an overdoped (oxygen-rich) 
crystal, it is expected to be positive \cite{chen_vortex_2002}. 
For a positively charged vortex core, the surface dipole moment points
in an outward direction from the BSCCO surface regardless to the 
direction of the applied magnetic field.
Therefore, the electromechanical response is expected
show an asymmetry with respect to the electric field
created by $V_{dc}$.
For a positively charged vortex core, the mechanical 
resonant frequency is expected 
to show a softening (hardening) behavior for negative (positive) 
$V_{dc}$ voltages as observed in the measurement.
Details of the electromechanical spring constant 
calculation are provided in the Supporting Information.
In next paragraph, we calculate the line charge density 
equivalent to this surface dipole moment and make a
comparison with the values reported in the literature. 

%%%
%%%

Following ref.~\cite{blatter_electrostatics_1996}, the charge density 
profile across a vortex core can be written as 
$\rho(r) \sim \left(ea_B\zeta/\pi^3\right)\left(\xi^2-r^2\right)/\left(\xi^2 + r^2\right)^3$, 
where $a_B$ is the Bohr radius, and $\zeta$ is the 
particle-hole asymmetry 
parameter. 
The line charge density in the vortex core, therefore, is given by
$Q_{\xi}~=\int_{0}^{\xi} \rho(r) 2\pi r \,dr = \zeta ea_B/\left(2\pi\xi\right)^2$. 
Such a charge redistribution is equivalent to a surface dipole pointing 
normal to the superconducting surface and having a magnitude of
$p_0=\left(\zeta ea_B/\pi^2\right)\left(m/m_{eff}\right)\ln\left(z/\xi\right)$.
Therefore, the line charge can be related to the dipole moment as
$Q_{\xi} = p_0 \left(m_{eff}/m\right) \left(4\xi^2 
\ln\left(z/\xi\right)\right)^{-1}$, where $m_{eff}/m$ is the effective mass
ratio of the charge carrier.
Using $\xi= 3.2$~nm 
\cite{naughton_orientational_1988},
$m_{eff}/m=4.6$ \cite{orlando_correlation_2018}, and $z=120$~nm, we obtain 
$Q_{\xi}=+4.9\times10^{-2}~|e|$/nm, which is
equivalent to a charge of $+3.7\times10^{-2}~|e|$ per CuO$_2$ layer.

Our estimation of the charge per copper oxide layer is consistent with 
the earlier measurements made on YBCO using the nuclear quadrupole 
resonance technique \cite{kumagai_charged_2001}. 
Both the experiments reveal a larger value of the vortex charge 
than that obtained via the theoretical predictions. 
The difference between the vortex charge magnitude from the 
experiments and the theory can have several interesting origins. 
First, it suggests that the gap anisotropy and Fermi surface 
curvature might be playing an important role \cite{ueki_vortex-core_2016}.
Second, given the ultralow temperature in this experiment, the quantum 
effects in the vortex could become relevant. Finally, the nature 
of the vortex core could be different from the metallic 
phase \cite{arovas_superconducting_1997, knapp_antiferromagnetism_2005}.

To summarize, we show clear evidence for the trapped-charges in
the vortex core probed using a cavity optomechanical device.
The magnitude of the equivalent dipole is higher than the
estimates from the BCS-theory while considering Thomas-Fermi screening.
Such a difference opens up new possibilities to revisit
the vortex-charge problem in atomically thin superconductors.
Furthermore, our experiment and novel device approach show the advantages 
of integrating exfoliated thin flake into the cavity optomechanical 
devices for sensitive measurement of the thermodynamical properties.
With minor modifications in the design, such technique could further 
be extended to other systems involving topological charge \cite{jiang_direct_2017}
or bosonic Landau levels \cite{devarakonda_signatures_2021}. 
This work thus opens up a new avenue to study quantum phase transitions as 
external parameters are varied.

\subsection{Acknowledgment}

The authors thank Eli Zeldov, Subir Sachdev 
and Vijay Shenoy for insightful discussions. The authors also
thank D.~Jangade and A. Thamizhavel for their help
during the crystal growth.
M.M.D. acknowledges the Department of Atomic Energy of the 
Government of India under Grant No. 12-R$\&$D-TFR-5.10-0100, 
DST Nanomission under Grant No. SR/NM/NS-45/2016 and  
SERB CORE grant CRG/2020/003836.
V.S. acknowledges the research support under the Core Research Grant
CRG/2018/001132 by DST and ISTC-0395 by STC-ISRO.
S.K.S. acknowledges device fabrication facilities at
the Department of Physics, and CeNSE,  IISc-Bangalore funded 
by Department of Science and Technology (DST), Government of India.

%\bibliographystyle{achemso}
%\bibliography{Ref}

\providecommand{\latin}[1]{#1}
\makeatletter
\providecommand{\doi}
{\begingroup\let\do\@makeother\dospecials
	\catcode`\{=1 \catcode`\}=2 \doi@aux}
\providecommand{\doi@aux}[1]{\endgroup\texttt{#1}}
\makeatother
\providecommand*\mcitethebibliography{\thebibliography}
\csname @ifundefined\endcsname{endmcitethebibliography}
{\let\endmcitethebibliography\endthebibliography}{}

\end{document}

\title{Superconducting vortex-charge measurement using cavity 
	electromechanics}

\author{Sudhir~Kumar~Sahu}
\affiliation{Department of Physics, Indian Institute of Science, 
	Bangalore-560012 (India)}

\author{Supriya~Mandal}
\affiliation{Department of condensed matter physics and material sciences, 
	Tata Institute of Fundamental Research, Mumbai - 400005 (India)} 

\author{Sanat~Ghosh}
\affiliation{Department of condensed matter physics and material sciences, 
	Tata Institute of Fundamental Research, Mumbai - 400005 (India)} 

\author{Mandar~M.~Deshmukh}
\affiliation{Department of condensed matter physics and material sciences, 
	Tata Institute of Fundamental Research, Mumbai - 400005 (India)}

\author{Vibhor~Singh}
\email[]{v.singh@iisc.ac.in}
\affiliation{Department of Physics, Indian Institute of Science, 
	Bangalore-560012 (India)}

\renewcommand{\thefigure}{S\arabic{figure}}
\renewcommand{\theequation}{S.\arabic{equation}}

\newcommand*\mycommand[1]{\texttt{\emph{#1}}}

\maketitle

\section{Measurement setup}

%%%%%%%%%%%%

\begin{figure*}
	\begin{center}
		\includegraphics[width = 120 mm]{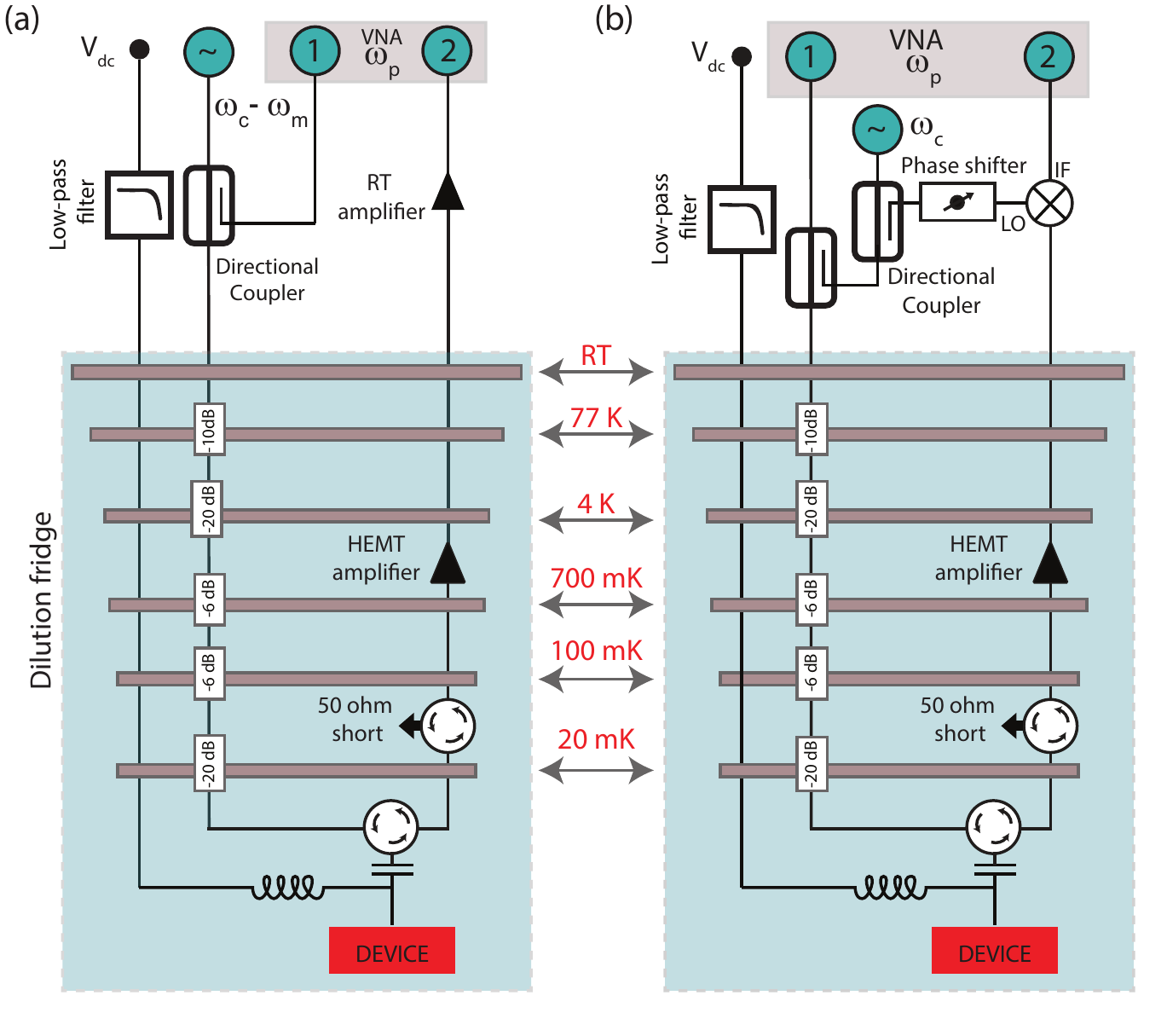}
		\caption{(a) Schematic of the measurement setup for the OMIA 
			measurements.
			(b) A schematic of the homodyne detection setup using a vector 
			network 
			analyzer (VNA). A mixer is used to demodulate the 
			reflected signal 
			from the cavity, and the intermediate frequency (IF) signal 
			is sent to the receiving port of the VNA.
			For measurement setup inside the dilution refrigerator, we use 
			Anritsu k250 bias tee, Pasternack PE8402 
			circulator and Raditek radc-4-8-Cryo-0.03-4K-S23-1WR-MS-b as 
			isolator.}
		\label{OMIA_setup}
	\end{center}
\end{figure*}

For microwave measurements, the device is mounted inside a microwave box 
and attached to the base plate of a dilution refrigerator at 20~mK. 
A coaxial cable is used as an input line with sufficient attenuators 
for microwave thermalization. 
The output signal is routed towards a HEMT amplifier 
using a coaxial line at 4~K stage followed by a microwave amplifier 
at room temperature.
A cryogenic circulator is used to measure the reflection from the device. 
Between the circulator and the device port, we use a bias-tee and a 
low-pass filter, which allows us to add a dc voltage to the
input feed-line of the cavity.
A directional coupler is used to combine microwave tones in the input lines at 
room temperature.
A detailed schematic of the measurement setup is given in 
Fig.~\ref{OMIA_setup}(a).

\subsection{Cavity characterization}

The microwave cavity is characterized by measuring the reflection 
coefficient $|S_{11}(\omega)|$ using a vector network analyzer (VNA). While characterizing the microwave cavity, the pump 
power at $\omega_c - \omega_m$ is kept off. 
To fit the reflection coefficient $|S_{11}(\omega)|$, we use

\begin{equation}
S_{11}(\omega)=\alpha e^{i\phi}+(1-\alpha)\left(1 - \frac{2\eta}{1+\frac{2i(\omega-\omega_c)}{\kappa}}\right),
\end{equation}
where $\alpha$ is the isolation of the circulator, $\eta$ is the ratio between 
the external dissipation rate $\kappa_e$ and the total dissipation 
rate $\kappa$, $\omega_c$ is the cavity resonant frequency and $\phi$ is 
a phase arising from the propagation delay.

\begin{figure*}
	\begin{center}
		\includegraphics[width = 140 mm]{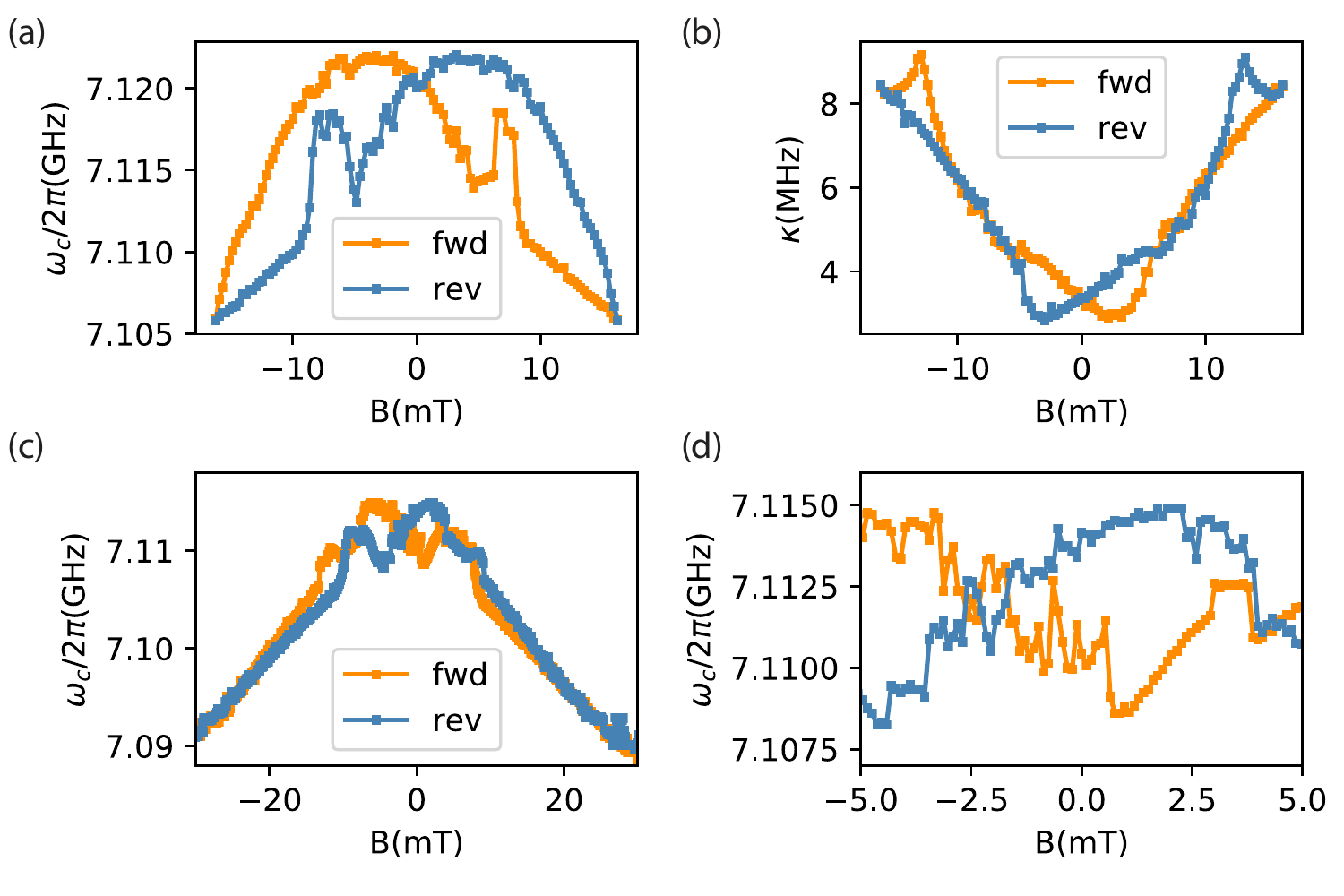}
		\caption{(a, b) For the sample studied in the main text, the change in 
			the resonant 
			frequency and linewidth of the cavity, respectively. The 
			magnetic field applied normal to 
			the cavity plane is swept back-and-forth. An abrupt change in 
			the cavity frequency indicates the movement of vortices in the
			film. 
			(c, d) Change in the resonant frequency of the bare microwave 
			cavity (a MoRe cavity without the BSCCO flake) as the magnetic 
			field applied normal to 
			the cavity plane is swept back-and-forth. The abrupt jumps are 
			found in cavity
			resonant frequency.}\label{cavity}
	\end{center}
\end{figure*}

For the characterization of the microwave cavity in the magnetic field,
the cavity resonant frequency and linewidth are measured with a step size of 
0.25~mT as shown in Fig.~\ref{cavity}(a) and Fig.~\ref{cavity}(b) respectively. 
It is interesting to point out that the maximum cavity frequency and 
the minimum cavity linewidth occur at different applied magnetic fields.
This is possibly due to the fact that the dominant contribution to the kinetic 
inductance and losses come from the different geometrical regions of
the coplanar cavity \cite{bothner_magnetic_2012}. While the kinetic inductance incorporates contributions 
from the vortices and the screening currents, the dissipation is dominated
by the dynamics taking place near the current anti-node (short boundary 
condition of the resonator).
Similar measurements of cavity resonant frequency 
were also carried out on a bare MoRe cavity as 
shown in Fig.~\ref{cavity}(c). We observe similar 
jumps in cavity frequency, which signifies that such
jumps are not unique to BSCCO based sample.
Fig.~\ref{cavity}(d) shows the 
zoomed-in view of the data set near the origin. The observation of jumps in the cavity frequency near zero magnetic field suggests the flux penetration at a field as small as 0.5~mT.

\subsection{Optomechanically-induced absorption measurements (OMIA)}
For mechanical mode characterization in an OMIA setup, we use a pump tone at 
the red sideband $(\omega_d = \omega_c - \omega_m)$ and a probe tone 
$(\omega_p)$ near cavity frequency. 
Both the microwave tones are added using a directional coupler as shown 
in Fig.~\ref{OMIA_setup}(a). As the magnetic field is swept, 
we first measure the cavity resonance as explained above at each set point. 
Subsequently, the red sideband pump tone and the cavity 
probe tone span are adjusted with the corresponding cavity frequency. 
The mechanical frequency is then calculated by fitting the OMIA 
response.

\subsection{Detection of mechanical mode using a low-frequency direct drive}

The mechanical mode can also be characterized independently using a 
direct-drive method. A low-frequency RF signal $V_{ac}$ and a dc
signal  $V_{dc}$ are sent to the feedline to `electrostatically’
apply a force $C'_gV_{ac}V_{dc}$ on the mechanical resonator, where $C'_g$ is 
the derivative of flake capacitance with respect to displacement.
Fig.~\ref{OMIA_setup}~(b) shows the detailed circuit diagram for the homodyne 
measurement.
In addition, a microwave tone at the resonant frequency of the cavity is 
added to the feedline. The motion of the mechanical resonator phase modulates 
the reflected signal from the cavity producing two sidebands at $\omega_c \pm 
\omega_{ac}$, where $\omega_{ac}$ is the frequency of the $ac$ drive.
The reflected signal is then demodulated using an external mixer at 
room temperature. This way, the cavity is effectively used as an interferometer.
The $ac$ drive and the demodulated signals are controlled and recorded 
by a vector network analyzer; thereby a measurement of $|S_{21}(\omega)|$ directly 
relates to the responsivity of the mechanical resonator.
Fig.~2(d) of the main text shows the colorplot of demodulated signal 
where a sharp change in color indicates the mechanical resonance.

\section{Estimation of the single-photon coupling rate}
The measurement of OMIA feature allows us to characterize the optomechanical 
interaction in terms of cooperativity $C = 4g^2/\kappa\gamma_m$, where $\kappa$ 
is the total cavity linewidth, $\gamma_m$ is the mechanical linewidth, 
$g=g_0\sqrt{n_d}$ is the linearized coupling strength and 
$n_d$ is the number of pump photons in the cavity. 
At $\omega_d = \omega_c -\omega_m$, the reflection the cavity frequency is given by 
$|S_{11}(\omega_c)|=\left|\frac{2\eta}{1+C}-1\right|$.
It thus
allows to estimate cooperativity directly from the swing
of the OMIA feature. 
Maximum cooperativity of 0.38 is achieved with 3~mW of red sideband pump
power at the signal generator.
Using the known attenuation in the setup, $n_d$ was estimated using $n_d = \frac{ P_{in} 
\kappa_e}{\hbar \omega_d \left(\left(\frac{\kappa}{2}\right)^2 + 
\left(\omega_d - \omega_c\right)^2\right)}$,
where $P_{in}$ is the power of the red sideband tone at the device input. 
Using the cooperativity results and the estimate of the number of photons,
we expect the single photon coupling rate to be $g_0/2\pi\approx30$~Hz.

\section{Estimation of the pre-tension in the mechanical mode}
The resonant frequency of the fundamental mode for a circular-shaped mechanical 
resonator clamped at the circumferences in the plate limit is given by 
\cite{timoshenko_vibration_2007}
\begin{equation*}
f_{p}= \frac{10.21}{\pi} \sqrt{\frac{E}{3\rho(1-\nu^2)}} \left(\frac{t}{d^2}\right),
\end{equation*}
where $t$ is the thickness, $d$ is the diameter, $E$ is the Young's modulus of 
rigidity, $\nu$ is the Poisson's 
ratio and $\rho$ is the volume mass density of the resonator. In the membrane 
limit, where pre-tension dominates over bending rigidity, the resonant 
frequency 
is given by \cite{timoshenko_vibration_2007}

\begin{equation*}
f_{t}=\frac{2.4048}{\pi d} \sqrt{\frac{T}{\rho t}}.
\end{equation*}

In the cross-over limit, both the contribution from plate limit and 
membrane limit are comparable. Hence the resonant frequency can be 
approximated to $f_m\approx\sqrt{f_{p}^2+f_{t}^2}$ 
\cite{castellanos-gomez_single-layer_2013}.
We measure the fundamental mode frequency of the BSCCO drum-head 
resonator to be 15.3826~MHz, a diameter of 7~$\mu$m 
and a thickness of 15~nm (5 unit cells). 
Using $E\sim$25~GPa, $\nu=$~0.2, and the mass density $\rho~=~2360~\text{Kg/m}^3$, 
we estimated the pre-tension $T_0$ to be 0.732~N/m \cite{sahu_elastic_2019}.

\section{Electromechanical response in the magnetic field}

To model the electromechanical response of the flake in the magnetic field, 
we consider the Lorentz force $F_L =2 \pi\int_{0}^{R} J(r) B_z(r) rdr$ 
acting on the BSCCO flake in the radial direction, where $J(r)$ is the 
screening current flowing in the circumferential direction and $B_z(r)$ is the $z$-component
of the local magnetic field.
In the critical state, the Lorentz force is balanced by the pinning forces. Therefore, it leads to a magnetic field-dependent stress in the suspended flake.
It can be obtained by $T_{L} = \int_{0}^{R} J(r)B_z(r) \,dr$.
To obtain the profiles $J(r)$ and $B_z(r)$, we use the 
exponential critical state model for thin flake and follow the 
treatment as prescribed by Shantsev \textit{et al.} \cite{shantsev_thin_1999}.

\subsection{Critical State Model}

The critical state models are widely used to explain the magnetic 
response of the high $T_c$ superconductors. Here we consider a 
thin disk of radius $R$ and thickness $t$ such 
that $\lambda ^2/t\ll R$, where $\lambda$ is the London penetration depth. 
This disk is placed in a perpendicular magnetic field pointing in the $z$-direction. 
Here the maximum sheet current $J_{c0}$ is defined as $J_{c0}=j_{c0} t$, where $j_{c0}$ 
is the bulk critical current density at zero applied field. 
With this consideration, the self-consistent equations for the sheet current 
$J(r)$, local magnetic field $B_z(r)$, and applied magnetic filed $B$ can be 
written as

\begin{equation}
    J(r) =
    \begin{cases}
    -\frac{2r}{\pi} \int_{a}^{R} \sqrt{\frac{a^2-r^2}{r'^2-a^2}} 
    \frac{J_c[B_z(r')]}{r'^2-r^2} \,dr' ,& \text{if } r < a\\
    -J_c[B_z(r)], & a<r<R
    \end{cases}
\end{equation}

\begin{equation}
B_z(r) = B + \frac{\mu_0}{2\pi} \int_{0}^{R} F(r,r') J(r') \,dr',
\end{equation}

\begin{equation}
B = B_c  \int_{a}^{R} \,dr'\frac{1}{\sqrt{r'^2-a^2}}\frac{J_c[B_z(r')]}{J_{c0}},
\end{equation}

where $a$ is the flux front position, $B_c = \mu_0 J_{c0}/2$  and 
$F(r,r')=K(u)/(r+r')-E(u)/(r-r')$, $u(r,r')= 2 \sqrt{r r'}/(r+r')$, and 
$K(u)$ and $E(u)$ are the 
complete elliptic integrals of second kind defined as $K(u) = \int_{0}^{\pi/2} 
[1-u^2 \cos^2{(x)}]^{-1/2} dx$ and $E(u) = \int_{0}^{\pi/2} [1-u^2 
\cos^2{(x)}]^{1/2} dx$.
For the exponential critical state model, we use 
$ J_c(B_z) = J_{c0} \exp{(-|B_z|/B_0)}$.

After increasing the field to a maximum value of $B_m$, 
the field and current profiles for the subsequent 
descent are calculated as follows:

Let us denote the flux front position, 
the current density, and the field distribution at the maximum 
field as $a_m$, $J_m$(r), and $B_{zm}$(r), respectively. 
During the magnetic field descent from $B_m$, the flux density becomes 
reduced in the outer region $a<r<R$, whereas the central part of 
the disk $r<a$ remains frozen.
During this process, local magnetic field $B_z(r)$ and the sheet current 
$J(r)$ given by

\begin{equation}\label{remag}
B_z(r)=B_{zm}(r)+\Tilde{B_z}(r), ~J(r)=J_m(r)+\Tilde{J}(r).
\end{equation}

For the calculation of $\Tilde{B_z}(r)$ and $\Tilde{J}(r)$ similar procedure is used with a modification of $J(r)=J_c[B_z(r)]$ in the region $a<r<R$.
The complete set of equations for field descent are given by 

\begin{equation}
    \Tilde{J}(r) =
    \begin{cases}
    \frac{2r}{\pi} \int_{a}^{R} \sqrt{\frac{a^2-r^2}{r'^2-a^2}} 
    \frac{\Tilde{J}_c[B_z(r')]}{r'^2-r^2} \,dr' ,& \text{if } r < a\\
    \Tilde{J}_c[B_z(r)], & a<r<R
    \end{cases}
\end{equation}

\begin{equation}
\Tilde{B}_z(r) = B-B_{m} + \frac{\mu_0}{2\pi} \int_{0}^{R} F(r,r') 
\Tilde{J}(r') \,dr',
\end{equation}

\begin{equation}
\frac{B-B_{m}}{B_c} = - \int_{a}^{R} \,dr\frac{1}{\sqrt{r^2-a^2}}\frac{\Tilde{J}_c[B_z(r)]}{J_{c0}},
\end{equation}

where $\Tilde{J}_c(r)= J_c[B_{zm}(r)+\Tilde{B_z}(r)]+J_c[B_{zm}(r)]$.

The above equations are solved numerically using the following iterative method. 
First, a flux front position is defined, which decides the initial guess for the applied magnetic field $B$. Further, $J_c$ is chosen to be independent of $B(r)$ for initial guess with a value of $J_{c0}$.
Equations are iterated with the initial guess until the solutions are converged.
With $B$ decreasing, the same procedure is used to find first
$J_m$(r), $B_{mr}$(r), $B_{m}$ for a given flux front position. Then, the 
equations for field descent are solved for a fixed flux front position yielding 
the functions $\Tilde{B_z}(r)$,
 $\Tilde{J}(r)$ and the applied field $B$.
Then the value of $B_z(r)$ and $J(r)$ for field descent state is calculated 
using (\ref{remag}). Once $B_z(r)$ and $J(r)$ are obtained, the Lorentz force 
and the additional tension $T_L$ in flake are calculated.
The mode frequency of the BSCCO resonator is then given by

 \begin{equation}
     f_m(B) = \sqrt{\left(\frac{2.4048}{\pi 
     d}\sqrt{\frac{T_0+\beta T_L(B) 
    }{\rho t} }\right)^2+ f_{plate}^2},
 \end{equation}
 
where $\beta$ is a geometrical factor included to account 
for the non-uniform shape of the flake. We mention 
the possible origin of such a geometrical factor in the next 
paragraph. For $\beta= 1/17=0.059$, the measured and the 
calculated mechanical frequency shift with a perpendicular 
magnetic field are shown in Fig.~3(a) of the main text. 

%%%%%%%%%%%%%%%%%%%%%%%%%%%%%%%%%%%%%%%%%%%%%%%%%%%%%%%%%%%%%%%%%%%%%%%%%%%%%

Due to the irregular shape of the entire BSCCO fake, 
the calculation of the screening 
current and the local magnetic field distribution is 
difficult. The critical state model discussed above 
assumes a circularly shaped BSCCO disk. Further, the 
non-uniform clamping of the suspended 
portion of the flake complicates the correction 
to the mechanical frequency shift due to the Lorentz 
force-induced stress. Therefore, the results obtained 
from the critical state model, assuming a circular geometry 
of the BSCCO flake, can not be directly applied.
By introducing the geometrical parameter $\beta$, we 
try to capture these aspects in a phenomenological 
manner.

\begin{figure}
\begin{center}
\includegraphics[width = 120 mm]{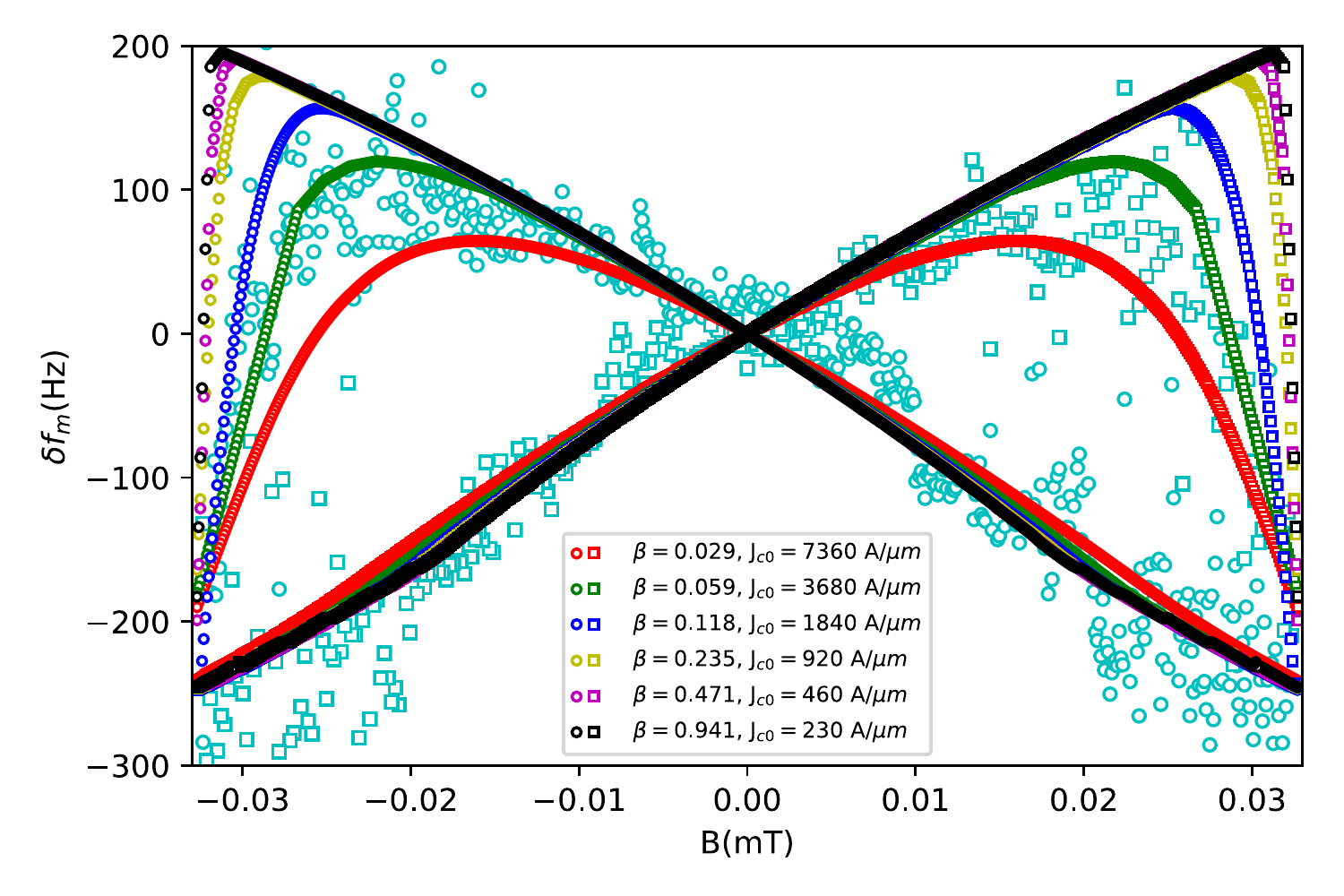}
\caption{Plot of the mechanical frequency shift with 
	the magnetic field for different combinations of 
	$J_{c0}$ and $\beta$. Data with circle and square 
	markers represent forward and reverse magnetic 
	sweep, respectively. 		
}
\label{beta_jc}
\end{center}
\end{figure}

To show the deviations, we have plotted the 
mechanical frequency shift with the magnetic field 
using different combinations of  $J_{c0}$ and $\beta$
in Fig.~\ref*{beta_jc}.
Circular (square) data points represent the frequency 
shift for the forward (reverse) sweep of the
magnetic field. For comparison, we have included 
the experimental data in cyan color. Variations in 
$\beta$ and $J_{c0}$ control the overall shape of 
hysteresis in the frequency shift. 
We find that the values of $\beta=0.059$
and $J_{c0}=~3680~$A/$\mu\text{m}$ produce results 
that resembles the experimental data. 
It is important to point out here that these choices
do not modify the final estimate of the vortex charge.

%%%%%%%%%%%%%%%%%%%%%%%%%%%%%%%%%%%%%%%%%%%%%%%%%%%%%%%%%%%%%%%%%%%%%%%%%%%%%
\section{Modification of the electromechanical response with $V_{dc}$}

To model the observed asymmetric change with respect to dc voltage, we include 
the energy of the effective surface dipoles formed by the flux-vortices \cite{blatter_electrostatics_1996}. 
When a dipole $\Vec{p}_0$ is placed in the electric field setup by the $V_{dc}$, 
it contributes $-\Vec{p}_0 \cdot \Vec{E}(z)$ to the electromechanical energy. Assuming a uniform distribution of the vortices in the suspended portion
of the BSCCO crystal, we can obtain their total contribution as
\begin{equation}
\begin{split}
U_{dip}(z) & = - n_v(B)\times \Vec{p}_0\cdot\Vec{E}(z) \\
              & \approx  n_v(B)\times p_0 \frac{V_{dc}}{z_0+z},
\end{split}
\end{equation}
where $n_v(B) = (\pi R^2 |B|)/\Phi_0$ is the total number of the vortices 
contained in the suspended portion of the flake, $z_0=120$~nm is the 
equilibrium separation between the ``plates" of the BSCCO capacitor 
with a capacitance $C_g$, and $\Phi_0$ is the superconducting
flux quantum.
We emphasis here that it is precisely the linear dependence
on $V_{dc}$ that results in the asymmetry in the response
with respect to the dc voltage.

With the dipole energy added, the total electromechanical energy of the 
drum can be written as,
\begin{equation}
\begin{split}
U(z) & = U_{elastic}+U_{electrostatic}+U_{dip} \\
& = \frac{1}{2} k_m z^2 - \frac{1}{2} C_g(z) V_{dc}^2 + n_v(B) 
p_0\frac{V_{dc}}{z_0+z},
\end{split}
\end{equation}

where $k_m = M\times(2\pi f_m(B,V_{dc}=0))^2$ is the mechanical spring constant, $M$
is the total mass of the resonator. With this total energy, the effective 
spring constant can be obtained as,

\begin{equation}
\begin{split}
k_{eff}(B, V_{dc}) & =\frac{\partial^2 U(z)}{\partial z^2}  \\
    & =  k_m - \frac{1}{2} \frac{\partial 
    ^2}{\partial z^2}C_g V_{dc}^2 + 2 n_v(B) p_0 
    \frac{V_{dc}}{z_0^3}.
\end{split}
\end{equation}

Results obtained from this equation are plotted in the Fig.~4 of the main text 
by defining, $f_{m}(B, V_{dc}) = \frac{1}{2\pi} \sqrt{\frac{k_{eff}}{M}}$.

%%%%%%%%%%%%%%%%%%%%%%%%%%%%%%%%%%%%%%%%%%%%%%%%%%%%%%%%%%%%%%%%%%%%%%%%%%%%%
\section{Optomechanical spring effect}

In a cavity optomechanical device, the measurement 
of mechanical resonant frequency can be affected 
due to the back action from the radiation-pressure force. 
The shift in the mechanical frequency while driving 
the system at red sideband is given by~\cite{aspelmeyer_cavity_2014}

\begin{equation*}
\delta f_m = g_0^2 \frac{P_{in}}{\hbar \omega_d} 
\frac{\kappa_e}{(\kappa^2/4)+\omega_m^2} 
\left(\frac{-2\omega_m}{(\kappa^2/4)+4\omega_m^2}\right),
\end{equation*}

where $P_{in}$ is the input drive power,
$g_0$ is the single photon coupling strength, and $\kappa = \kappa_i + 
\kappa_e$ is the total dissipation rate of the microwave cavity.
A constant shift in the mechanical frequency does not affect 
the results obtained in the magnetic field. However, due to the magnetic field 
dependent loss rate of the cavity, the circulating energy in the
cavity ($\hbar\omega_d n_d$) changes. It thus leads to a shift in the
mechanical resonant frequency.
As our device operates in the resolved-sideband limit 
$(\kappa < \omega_m)$, the correction
to the mechanical frequency due to change in cavity 
linewidth is very small. Using estimated injected power 
of -61~dBm and a variation in $\kappa$ from 4~MHz to 
9~MHz, we calculated the worst-case shift in the 
mechanical frequency to be $\sim$~8~Hz.

%%%%%%%%%%%%%%%%%%%%%%%%%%%%%%%%%%%%%%%%%%%%%%%%%%%%%%%%%%%%%%%%%%%%%%%%%%%%%
\section{Screening of the magnetic field by a superconducting disk}

\begin{figure}
	\begin{center}
		\includegraphics[width = 120 mm]{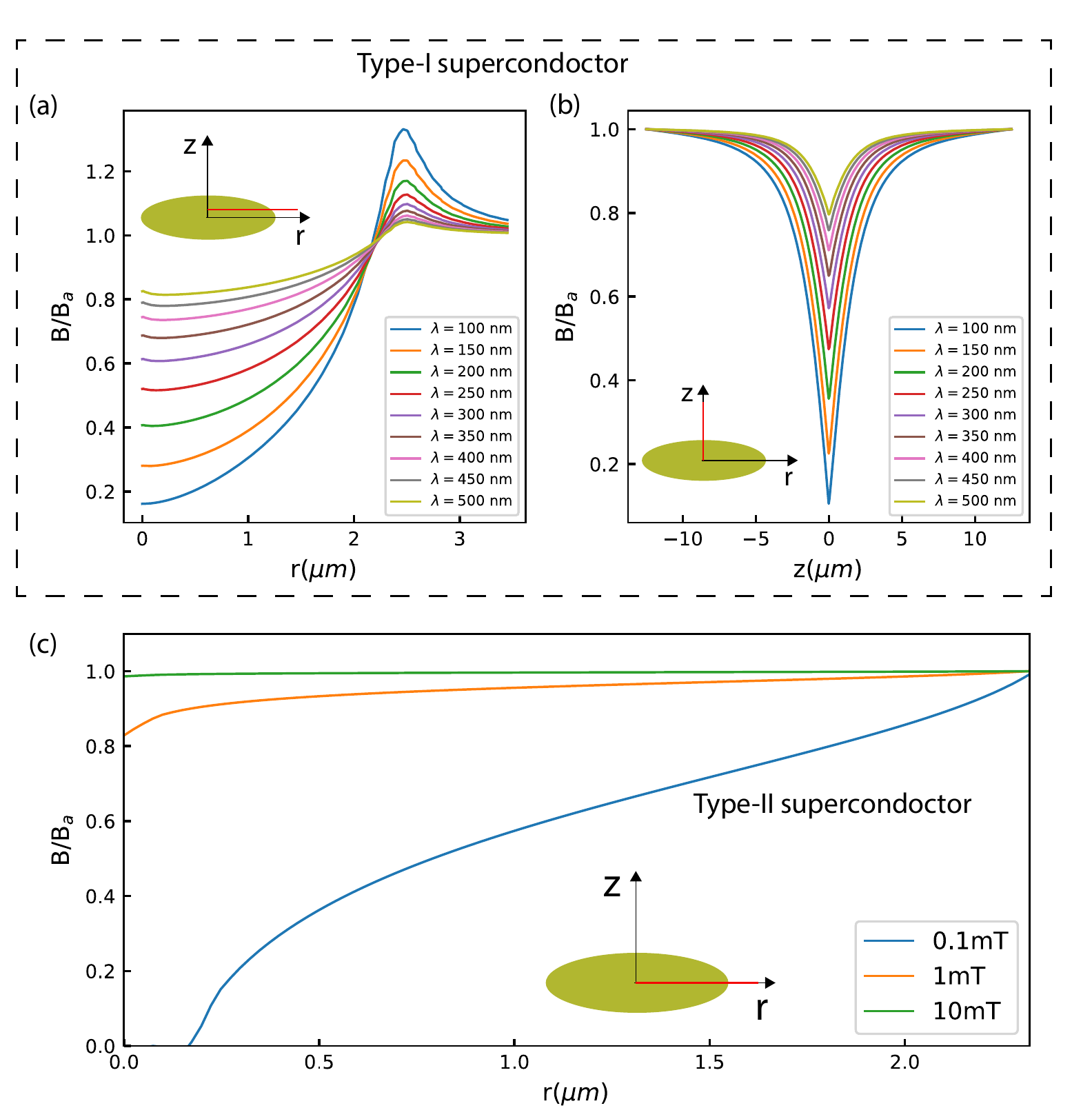}
		\caption{Screening by a superconducting disk
			in an applied magnetic field $B_a$ along the
			axis of the disk: Results obtained from FEM 
			simulations assuming a type-I superconductor 
			for different values of penetration length. 
			(a) Plot of the normalized magnetic field the 
			center of the disk along the radial direction 
			in a plane 120~nm above the disk. The inset 
			shows the schematic of the disk and its orientation. 
			(b) Plot of the normalized magnetic field along 
			the axis of the disk.
			(c) Results from the critical state model for a 
			type-II superconductor: Normalized field at the 
			surface of MoRe disk plotted from the center of 
			the disk along the radial direction. The radius 
			of the disk is taken as 2.5~$\mu$m. }
		\label{screening}
	\end{center}
\end{figure}

To understand the magnetic field screening by the MoRe electrode, 
positioned underneath the BSCCO flake, we take two approaches. 
First, we treat MoRe as a type-I superconductor having  
the London penetration length $\lambda$. 
We write down the governing equations
for the vector potential as described by 
Caputo et al. \cite{caputo_screening_2013}.
Using the finite-element techniques, we numerically 
solve these equations to obtain the magnetic field at 
the position of the BSCCO flake (\textit{i.e.} 120 nm above the 
MoRe electrode). We consider a $5~\mu$m diameter 
and 80~nm thick circular superconducting disk 
placed in a magnetic field parallel to the 
axis of the disk.
Fig.~\ref{screening}~(a) shows the plot of the 
normalized magnetic field strength at different 
radial distances in a plane 120~nm above the disk.
In addition, Fig.~\ref{screening}~(b) shows 
the plot of the normalized magnetic field along 
the axis of the disk. 
We see that as the penetration length is increased,
the magnetic field screening by the bottom 
disk deteriorates. 
For $\lambda =  500$~nm, nearly 75~$\%$ of the 
magnetic field is recovered at the location 
of the BSCCO flake.

Since MoRe is a type-II superconductor, we use 
the critical state model to compute the normal 
component of the magnetic field at the surface of
the bottom electrode. We use a critical current 
density of $J_{c0} = 6.2\times 10^8 $~A/m$^2\times80$~nm 
\cite{sundar_electrical_2013}, and $B_0 = 100$~mT.
The calculated local magnetic field 
for 3 different values of applied magnetic field is 
plotted in Fig.~\ref*{screening}~(c).
These results suggest that the MoRe bottom electrode 
offers negligible screening for the perpendicular 
magnetic field.

%%%%%%%%%%%%%%%%%%%%%%%%%%%%%%%%%%%%%%%%%%%%%%%%%%%%%%%%%%%%%%%%%%%%%%%%%%%%%

%\newpage{}
\begin{table}
	\caption{Device parameter table}
	\begin{tabular}{|c|c|}
		\hline
		Parameter & Value \\
		\hline
		$g_0/2\pi$ & 30~Hz   \\
		\hline
		$\omega_c/2\pi(V_{dc}=0,B=0)$ & 7.124~GHz   \\
		\hline
		$\omega_m/2\pi(V_{dc}=0,B=0)$ & 15.383~MHz  \\
		\hline
		typical probe photons & 0.15$\times 10^6$   \\
		\hline
		typical drive photons & 0.55$\times 10^6$   \\
		\hline
		total mass of resonator $(M)$ & 1.3$\times 10^{-15}$~kg   \\
		\hline
		quantum zero point motion $\left(x_{zp}\right)$ & 20~fm\\
		\hline
		BSCCO drum capacitance $c_g$ & 2.1~fF \\
		\hline
		coherence length of BSCCO $(\xi)$ & 3.2~nm \\
		\hline
		effective mass ratio of charge carrier $m_{eff}/m$ & 4.6 \\
		\hline
	\end{tabular}\label{tbl1}
\end{table}

\newpage{}
%\bibliographystyle{achemso}
%\bibliography{Refs_SI}

\providecommand{\latin}[1]{#1}
\makeatletter
\providecommand{\doi}
{\begingroup\let\do\@makeother\dospecials
	\catcode`\{=1 \catcode`\}=2 \doi@aux}
\providecommand{\doi@aux}[1]{\endgroup\texttt{#1}}
\makeatother
\providecommand*\mcitethebibliography{\thebibliography}
\csname @ifundefined\endcsname{endmcitethebibliography}
{\let\endmcitethebibliography\endthebibliography}{}